\documentclass[11pt,tightenlines,superscriptaddress,floatfixolumn,english,aps,
notitlepage,nofootinbib]{revtex4-1}
\usepackage{amsmath}
\usepackage{esint}
\usepackage{graphicx}
\usepackage{hyperref}
\usepackage{color}
\usepackage{xcolor}
\usepackage[utf8]{inputenc} %%% needed  for some of the references
\usepackage{enumitem}
\usepackage{float}
\usepackage{array}
\usepackage[norefs,nocites]{refcheck}	%%% checking for unused refrence
\usepackage{subfig}
\usepackage{soul}
\usepackage[makeroom]{cancel}

\advance\voffset -0.2in

\newcommand\D{\text{d}}
\newcommand\E{\text{e}}

\newcommand\ds{\displaystyle}

\newcommand\HUGE{\@setfontsize\Huge{38}{47}} 
\newcolumntype{C}{>{\centering\arraybackslash}p{5em}}

\newcolumntype{R}[1]{>{\raggedleft\let\newline\\\arraybackslash\hspace{0pt}}m{#1}}

\definecolor{darkgreen}{HTML}{427F48}

\begin{document}
%\norefnames

%\preprint{}
\title{
Wounded nucleon, quark and quark-diquark emission functions versus experimental results from RHIC 
}

\author{Micha{\l} Barej}
\email{michal.barej@fis.agh.edu.pl}
\affiliation{AGH University of Science and Technology,
Faculty of Physics and Applied Computer Science,
30-059 Krak\'ow, Poland}

\author{Adam Bzdak}
\email{adam.bzdak@fis.agh.edu.pl}
\affiliation{AGH University of Science and Technology,
Faculty of Physics and Applied Computer Science,
30-059 Krak\'ow, Poland}

\author{Pawe{\l} Gutowski}
\email{pawel.gutowski@fis.agh.edu.pl}
\affiliation{AGH University of Science and Technology,
Faculty of Physics and Applied Computer Science,
30-059 Krak\'ow, Poland}

%%%%%%%%%%%%%%%%%%%%%%%%%%%%%%%%%%%%%%%%%%%%%%%%%%%%%%%%%%%%%%%%%%%%%%%%%%%%%%%%%%%%%%%%%%
%%%%%%%%%%%%%%%%%%%%%%%%%%%%%%%%%%%%%%%%%%%%%%%%%%%%%%%%%%%%%%%%%%%%%%%%%%%%%%%%%%%%%%%%%%
%%%%%%%%%%%%%%%%%%%%%%%%%%%%%%%%%%%%%%%%%%%%%%%%%%%%%%%%%%%%%%%%%%%%%%%%%%%%%%%%%%%%%%%%%%

\begin{abstract}

Using the wounded nucleon, quark, and quark-diquark models, we extract the wounded source emission functions from the PHOBOS data for d+Au collisions at $\sqrt{s_{_{NN}}}=200$ GeV. We apply these models to compute charged particle multiplicity distributions as functions of pseudorapidity for p+p, p+Al, p+Au, d+Au, $^3$He+Au, Cu+Cu, Cu+Au, Au+Au, and U+U collisions at the same energy and compare them with experimental data from the PHOBOS and PHENIX collaborations. In symmetric collisions of heavy nuclei, the obtained distributions differ among the tested models. On the other hand, in asymmetric collisions, all three models give essentially the same distributions. The wounded quark-diquark and quark models are in reasonable agreement with data for all the investigated systems.

\end{abstract}

\maketitle

%%%%%%%%%%%%%%%%%%%%%%%%%%%%%%%%%%%%%%%%%%%%%%%%%%%%%%%%%%%%%%%%%%%%%%%%%%%%%%%%%%%%%%%%%%
%%%%%%%%%%%%%%%%%%%%%%%%%%%%%%%%%%%%%%%%%%%%%%%%%%%%%%%%%%%%%%%%%%%%%%%%%%%%%%%%%%%%%%%%%%
%%%%%%%%%%%%%%%%%%%%%%%%%%%%%%%%%%%%%%%%%%%%%%%%%%%%%%%%%%%%%%%%%%%%%%%%%%%%%%%%%%%%%%%%%%

% ${\D N_{\text{ch}}}/{\D \eta}$

\section{Introduction}
In this paper, we examine various wounded source models commonly used to characterize the particle production in ultrarelativistic heavy-ion collisions.
In the wounded nucleon model \cite{Bialas:1976ed}, a heavy ion collision is described as a superposition of multiple independent nucleon-nucleon interactions. The wounded quark model \cite{Bialas:1977en} is similar but constituent quark-quark collisions are considered. This model was studied recently in various contexts with rather interesting results \cite{Adler:2013aqf,Adare:2015bua,Bozek:2016kpf,Lacey:2016hqy,Loizides:2016djv,Mitchell:2016jio,Bozek:2017elk,Chaturvedi:2016ctn,Zheng:2016nxx,Barej:2017kcw,Rohrmoser:2018shp}. In the wounded quark-diquark model \cite{abab}, a nucleon is assumed to consist of a quark and a diquark. This model not only reproduces charge particle multiplicities but also naturally explains the differential elastic cross-section in proton-proton collisions in a broad range of energies \cite{abab,Bzdak:2007qq,Nemes:2012cp,Csorgo:2013bwa,CsorgO:2013kua,Nemes:2015iia}. In all three wounded source models, it is assumed that every constituent (nucleon, quark or diquark, depending on the model) which underwent at least one inelastic collision produces particles independently of the number of collisions. Such sources we call wounded. 

In our previous work \cite{Barej:2017kcw}, we investigated the wounded quark emission function $F(\eta)$ extracted from d+$^{197}\!$Au collisions at $\sqrt{s_{_{NN}}}=200\ \text{GeV}$ measured by the PHOBOS collaboration at the Relativistic Heavy Ion Collider (RHIC) \cite{Back:2004mr}. This function is defined as the pseudorapidity particle multiplicity distribution from one wounded source. We observed that the wounded-quark emission function, $F(\eta)$, is practically universal in various centrality classes, see also Ref. \cite{Rohrmoser:2018shp}. Using min-bias $F(\eta)$, we predicted the pseudorapidity charged particle multiplicity distributions, ${\D N_{\text{ch}}}/{\D \eta}$, for p+$^{27}\!$Al, p+$^{197}\!$Au and $^3$He+$^{197}\!$Au. Our predictions turned out to be in good agreement with recent experimental results from the PHENIX collaboration \cite{Adare:2018toe}. 

In this paper, we study three different emission functions extracted from the wounded nucleon model (WNM), the wounded quark model (WQM) and the wounded quark-diquark model (WQDM). Using these functions we compute ${\D N_{\text{ch}}}/{\D \eta}$ distributions in various centrality classes at $\sqrt{s_{_{NN}}}=200$ GeV for several colliding systems measured by PHOBOS and PHENIX. Our goal is to determine which model is best to reproduce the experimental results.

As expected, for symmetric collisions such as $^{63}$Cu+$^{63}$Cu or $^{197}\!$Au+$^{197}\!$Au there are significant differences among the studied models.
Namely, the wounded nucleon model is unsuitable for these collisions (except very peripheral ones), whereas both the wounded quark and quark-diquark models are in good agreement with the RHIC results on ${\D N_{\text{ch}}}/{\D \eta}$ \cite{Alver:2007aa,Back:2002wb}.
On the other hand, in asymmetric collisions (such as the ones studied in, e.g., 
Ref. \cite{Adare:2018toe}) with one light nucleus 
(p, d and $^{3}$He) all three models give practically identical results. We also made simulations for $^{63}$Cu+$^{197}\!$Au and $^{238}$U+$^{238}$U and compared them with available data from PHENIX \cite{Adare:2015bua}.

This paper is organized as follows. In the next section, there is a brief description of the wounded source models. Next, we show the minimum bias emission functions extracted from the PHOBOS data for d+$^{197}\!$Au collisions. Finally, we compare our simulations with the RHIC results for nucleus-nucleus collisions at $\sqrt{s_{_{NN}}}=200\ \text{GeV}$. In the last section, our conclusions are presented.

%%%%%%%%%%%%%%%%%%%%%%%%%%%%%%%%%%%%%%%%%%%%%%%%%%%%%%%%%%%%%%%%%%%%%%%%%%%%%%%%%%%%%%%%%%
%%%%%%%%%%%%%%%%%%%%%%%%%%%%%%%%%%%%%%%%%%%%%%%%%%%%%%%%%%%%%%%%%%%%%%%%%%%%%%%%%%%%%%%%%%
%%%%%%%%%%%%%%%%%%%%%%%%%%%%%%%%%%%%%%%%%%%%%%%%%%%%%%%%%%%%%%%%%%%%%%%%%%%%%%%%%%%%%%%%%%

\section{Wounded source models}
 
We consider three models: the wounded nucleon model (WNM), the wounded quark-diquark model (WQDM) and the wounded quark model (WQM).
The models differ by the composition of a nucleus, i.e., a nucleus consists of nucleons in the WNM (no internal structure), constituent quarks and diquarks in the WQDM, and constituent quarks in the WQM. In all three models, we assume that every wounded constituent populates charged particles regardless of the number of collisions \cite{Bialas:1976ed}. Therefore, we can treat a collision of two nuclei as a superposition of independent nucleon-nucleon (WNM), quark-quark (WQM) and quark(diquark)-quark(diquark) (WQDM) interactions. In general, the single particle distribution of charged particles is given by \cite{Bialas:2004su}
\begin{equation}\label{eq:start}
\frac{\D N_{\text{ch}}}{\D \eta}=w_LF(\eta)+w_RF(-\eta)\,,
\end{equation}
where $F(\pm\eta)$ is the emission function of one constituent (nucleon, quark or diquark) and the coefficients $w_L,\,w_R$ are the mean numbers of wounded constituents in the left-going and the right-going nuclei, respectively. If $w_L\neq w_R$ we can extract $F(\eta)$
\begin{equation}\label{eq0}
F(\eta)=\frac{1}{2}\bigg[\frac{N(\eta)+N(-\eta)}{w_L+w_R}+\frac{N(\eta)-N(-\eta)}{w_L-w_R}\bigg]\,,
\end{equation}
where $N(\pm\eta)={\D N_{\text{ch}}(\pm\eta)}\big/{\D \eta}$. The mean numbers of wounded sources are calculated using the Glauber Monte Carlo simulations with parameters listed in Ref. \cite{Loizides:2014vua}.

The positions of nucleons in nuclei are randomly drawn from the appropriate distributions. For deuteron, the Hulthen formula determines the proton's position
\begin{equation}
\rho(\vec{r})=\rho_0\bigg(\frac{\E^{-Ar}+\E^{-Br}}{r}\bigg)^2\,,
\end{equation}
where $r$ is a distance from the center of a nucleon with parameters $A=0.457$ fm$^{-1}$, $B=2.35$ fm$^{-1}$ and the neutron is placed opposite to the proton \cite{hulthen,Loizides:2014vua}.
For helium-3 we used \cite{Carlson:1997qn} to determine the nucleons' coordinates $\vec{r}$.
For gold, copper, aluminium and uranium, the positions of nucleons are given by the Woods-Saxon distribution
\begin{equation}
\rho(\vec{r})=\rho_0\Bigg({1+\exp\bigg(\frac{r-R(1+\beta_2Y_{20}+\beta_4Y_{40})}{a}\bigg)}\Bigg)^{-1}\,,
\label{eq:WS}
\end{equation}
where $Y_{20}=\sqrt{ 5\over{16\pi}}(3\cos^{2}\theta-1)$, $Y_{40}={ 3\over {16\sqrt{\pi}}}(35\cos^{4}\theta-30\cos^{2}\theta+3)$ and all the parameters are listed in Table \ref{table0}  \cite{DeJager:1987qc,Loizides:2014vua}.
%%%%%%%%%%%%%%%%%%%%%%%%%%%%%%%%%%%%%%%%%%%%%%%%%%%%%%%%%%%%%%%%%%%%%%%%%%%%%%%%%%%%%%%%%%
\begin{table}[H]\centering
\begin{tabular}{|*{5}{C|}} \hline
 & $a~[\text{fm}]$ & $R~[\text{fm}]$ & $\beta_2$ & $\beta_4$ \\ \hline
$^{27 }\!$Al     & 0.580 & 3.34 & -0.448 & 0.239 \\ \hline
$^{63 }  $Cu     & 0.596 & 4.20 &  0     &  0    \\ \hline
$^{197}\!$Au     & 0.535 & 6.38 &  0     &  0    \\ \hline
$^{238}  $U$~~$  & 0.440 & 6.67 & ~0.280 & 0.093 \\ \hline
\end{tabular}
\caption{\label{table0} Parameters used in our calculations for the Woods-Saxon distribution, see Eq. (\ref{eq:WS}).}
\end{table}
%%%%%%%%%%%%%%%%%%%%%%%%%%%%%%%%%%%%%%%%%%%%%%%%%%%%%%%%%%%%%%%%%%%%%%%%%%%%%%%%%%%%%%%%%%
For the WQM we generate three quarks independently according to \cite{Hofstadter:1956qs}
\begin{equation}\label{eq1}
\rho(\vec{r})=\rho_0\exp\left(-\sqrt{12} C r/ r_p\right)\,,
\end{equation} 
with $r_p=0.81\,\text{fm}$ being the proton's radius and the coefficient $C=0.82$ results from shifting the quarks to the center-of-mass of a nucleon.\footnote{We choose the parameter $C$ in Eq. (\ref{eq1}) so that $\langle r^2 \rangle = r_{p}^2$ for generated quarks or quarks and diquarks.}
In the WQDM, we generate a quark at a distance $r$ from the center of mass according to Eq. (\ref{eq1}) with $C=0.79$ and then place a diquark in the opposite direction at a distance of $r/2$. This is, of course, equivalent to the assumption that a diquark is two times heavier than a quark. We verified that this assumption is not crucial in our calculations. Assuming for example that both masses are equal, we obtained almost identical results.

In the next step, we draw the squared impact parameter $b^2$ from a uniform distribution in an interval of [0,$b_{\text{max}}^{\,2}$]. We took $b_{\text{max}}$ to be 5 fm for p+p, 9 fm for p+Al, 15 fm for p+Au, d+Au, $^3$He+Au, Cu+Cu, 18 fm for Cu+Au, Au+Au, and 20 fm for U+U collisions. Then we count wounded sources by checking whether each source from one nucleus collided with at least one source from another one.
To determine if two constituents interact with each other we used a normal distribution  
and checked whether the transverse distance $s$ between colliding sources and the random variable $u$ (from a uniform distribution in $[0,1]$) satisfy $\ds u<\exp(-s^2\,\big/\,2\gamma^2)$, where $\gamma^2=\sigma_{\text{ii}}\,/\,2\pi$ and $\sigma_{\text{ii}}$ is an inelastic constituent-constituent cross-section.
For collisions at $\sqrt{s_{_{NN}}}=200\,\text{GeV}$ we took the corresponding nucleon-nucleon cross-section of $\sigma_{\text{nn}} = 41$ mb \cite{Loizides:2014vua}. For the WQM we determined $\sigma_{\text{qq}}$ using the trial and error method \cite{Barej:2017kcw}. We found the value of 
$\sigma_{\text{qq}}=6.65\,\text{mb}$, which satisfies $ \sigma_{\text{nn}} = \int_{0}^{2 \pi} d\varphi \int_{0}^{+ \infty}\!ds\,s P(s)$, where $P(s)$ is a probability for the inelastic collision of two nucleons with the transverse distance $s$. In the WQDM there are three possible types of collisions of constituents: quark-quark, quark-diquark and diquark-diquark. With this in mind, we modified our trial and error procedure for the WQDM assuming that the corresponding cross-sections satisfy the proportion $\sigma'_{\text{qq}}:\sigma_{\text{qd}}:\sigma_{\text{dd}}=1:2:4$ \cite{abab}. Using the trial and error method we found that $\sigma'_{\text{qq}}=5.75\,\text{mb}$. 

To calculate the number of emitted charged particles (used, e.g., to determine the centrality class) we assumed that each wounded source populates particles according to a negative binomial distribution, described by two parameters $\langle n\rangle$ and $k$, the latter characterizing the deviation from a Poisson distribution.
In the WNM we took $\langle n\rangle=5$ and $k=1$ \cite{Ansorge:1988kn}. For the WQM and the WQDM these numbers should be divided by the mean numbers of wounded constituents in a nucleon-nucleon collision, which are equal to $1.27$ (WQM) and $1.14$ (WQDM) per one nucleon.

%%%%%%%%%%%%%%%%%%%%%%%%%%%%%%%%%%%%%%%%%%%%%%%%%%%%%%%%%%%%%%%%%%%%%%%%%%%%%%%%%%%%%%%%%%
%%%%%%%%%%%%%%%%%%%%%%%%%%%%%%%%%%%%%%%%%%%%%%%%%%%%%%%%%%%%%%%%%%%%%%%%%%%%%%%%%%%%%%%%%%
%%%%%%%%%%%%%%%%%%%%%%%%%%%%%%%%%%%%%%%%%%%%%%%%%%%%%%%%%%%%%%%%%%%%%%%%%%%%%%%%%%%%%%%%%%

\section{Emission functions}

Our goal is to extract the wounded source emission functions from the minimum-bias d+Au collisions measured by PHOBOS. Next, using Eq. (\ref{eq:start}) we can calculate ${\D N_{\text{ch}}}/{\D \eta}$ for all colliding systems and all centralities of interest. We note that this procedure does not introduce any free parameters (the only parameters we use are the ones from the Woods-Saxon distribution etc.)

\begin{figure}[H]
\begin{center}
\subfloat{{\includegraphics[scale=0.30]{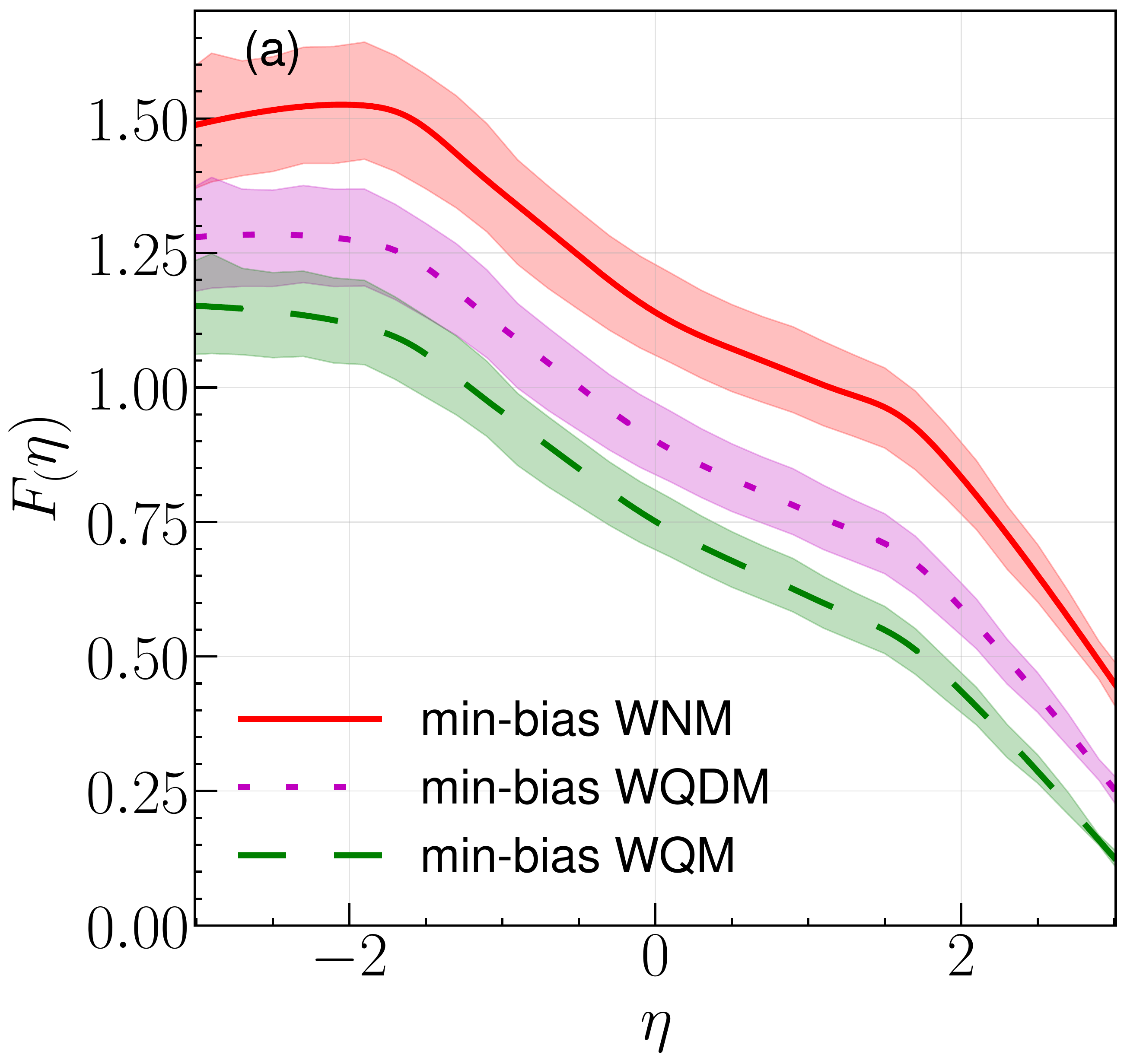}}}%
\hspace{0.3cm}
\subfloat{{\includegraphics[scale=0.30]{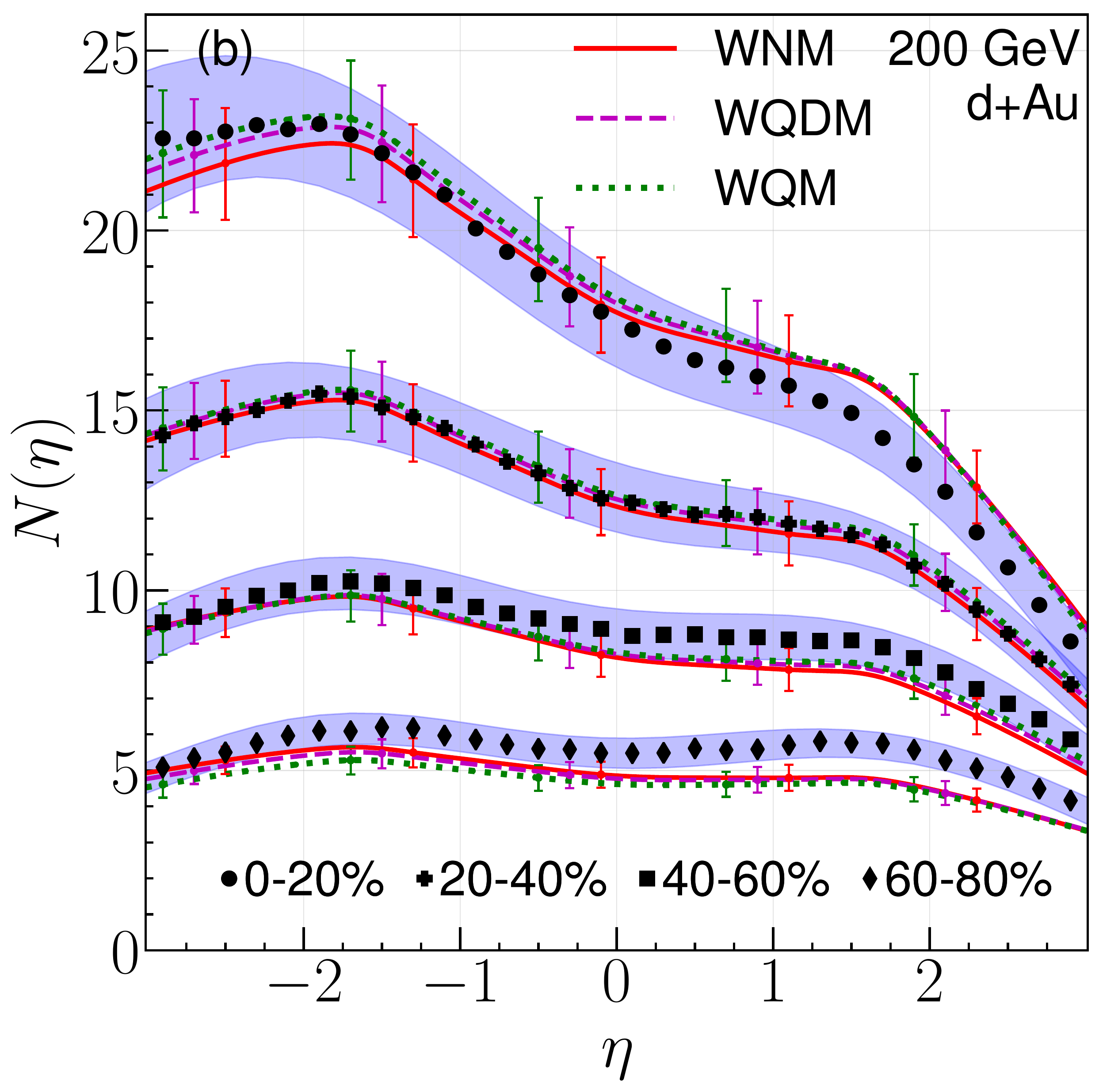}}}%
\caption{(a) The min-bias wounded nucleon, quark-diquark, and quark emission functions at $\sqrt{s_{_{NN}}}=200$ GeV. (b) Reconstruction of charged particle multiplicity distributions $N(\eta)\equiv\D N/\D\eta$ as functions of pseudorapidity for d+$^{197\!}$Au using the calculated min-bias emission functions. Points represent the PHOBOS data and lines represent our simulation results. Shaded areas and bars represent corresponding uncertainties for the PHOBOS data and our calculations, respectively.}\label{fig:F0}
\end{center}
\end{figure}

Performing the Glauber Monte-Carlo calculations, described in the previous Section, we determined the mean numbers of wounded constituents in min-bias d+$^{197\!}$Au collisions, see Table \ref{table:d-Au0} (where we also show the results for other measured centralities). With these values and the PHOBOS data, the min-bias emission functions for wounded nucleons, quarks, and diquarks are calculated according to Eq. (\ref{eq0}) and presented in Figure \ref{fig:F0}(a). We note that in this calculation we fitted the PHOBOS data with an analytical function, as described in the appendix A.
%%%%%%%%%%%%%%%%%%%%%%%%%%%%%%%%%%%%%%%%%%%%%%%%%%%%%%%%%%%%%%%%%%%%%%%%%%%%%%%%%%%%%%%%%%
\begin{table}[H]\centering
\begin{tabular}{|R{1.5cm}|R{2cm}|R{2cm}|R{2cm}|R{2cm}|R{2cm}|} \hline
 & min-bias & 0-20\% & 20-40\% & 40-60\% & 60-80\% \\ \hline
WNM  & 1.60, 6.56 & 1.97, 13.59 & 1.86,  8.96 & 1.65, 5.49 & 1.36, 2.90 \\ \hline
WQDM & 2.65, 7.67 & 3.80, 16.15 & 3.39, 10.52 & 2.74, 6.40 & 1.95, 3.33 \\ \hline
WQM  & 3.68, 8.70 & 5.63, 18.48 & 4.89, 11.94 & 3.78, 7.25 & 2.49, 3.66 \\ \hline
\end{tabular}
\caption{The mean numbers of wounded nucleons, quark-diquarks and quarks for various centrality classes in d+$^{197\!}$Au collisions at $\sqrt{s_{_{NN}}} = 200\  \text{GeV}$. The first and second numbers in each cell concern d and Au, respectively.}\label{table:d-Au0}
\end{table}
%%%%%%%%%%%%%%%%%%%%%%%%%%%%%%%%%%%%%%%%%%%%%%%%%%%%%%%%%%%%%%%%%%%%%%%%%%%%%%%%%%%%%%%%%%
%%%%%%%%%%%%%%%%%%%%%%%%%%%%%%%%%%%%%%%%%%%%%%%%%%%%%%%%%%%%%%%%%%%%%%%%%%%%%%%%%%%%%%%%%%
In Fig. \ref{fig:F0}(b) we show how our models reproduce the data for d+$^{197}$Au at $\sqrt{s_{_{NN}}}=200$ GeV, based on the min-bias emission functions.\footnote{In Fig. \ref{fig:F0}(b) we do not show the min-bias points since they are described in our models by construction.} One can see that all models describe the data quite well, although certain discrepancies can be observed. We also note that all models give practically identical results.

%%%%%%%%%%%%%%%%%%%%%%%%%%%%%%%%%%%%%%%%%%%%%%%%%%%%%%%%%%%%%%%%%%%%%%%%%%%%%%%%%%%%%%%%%%
%%%%%%%%%%%%%%%%%%%%%%%%%%%%%%%%%%%%%%%%%%%%%%%%%%%%%%%%%%%%%%%%%%%%%%%%%%%%%%%%%%%%%%%%%%
%%%%%%%%%%%%%%%%%%%%%%%%%%%%%%%%%%%%%%%%%%%%%%%%%%%%%%%%%%%%%%%%%%%%%%%%%%%%%%%%%%%%%%%%%%

\section{Results}
\subsection{p+Al, p+Au, d+Au and He+Au}
\begin{figure}[H]
\begin{center}
\includegraphics[scale=0.20]{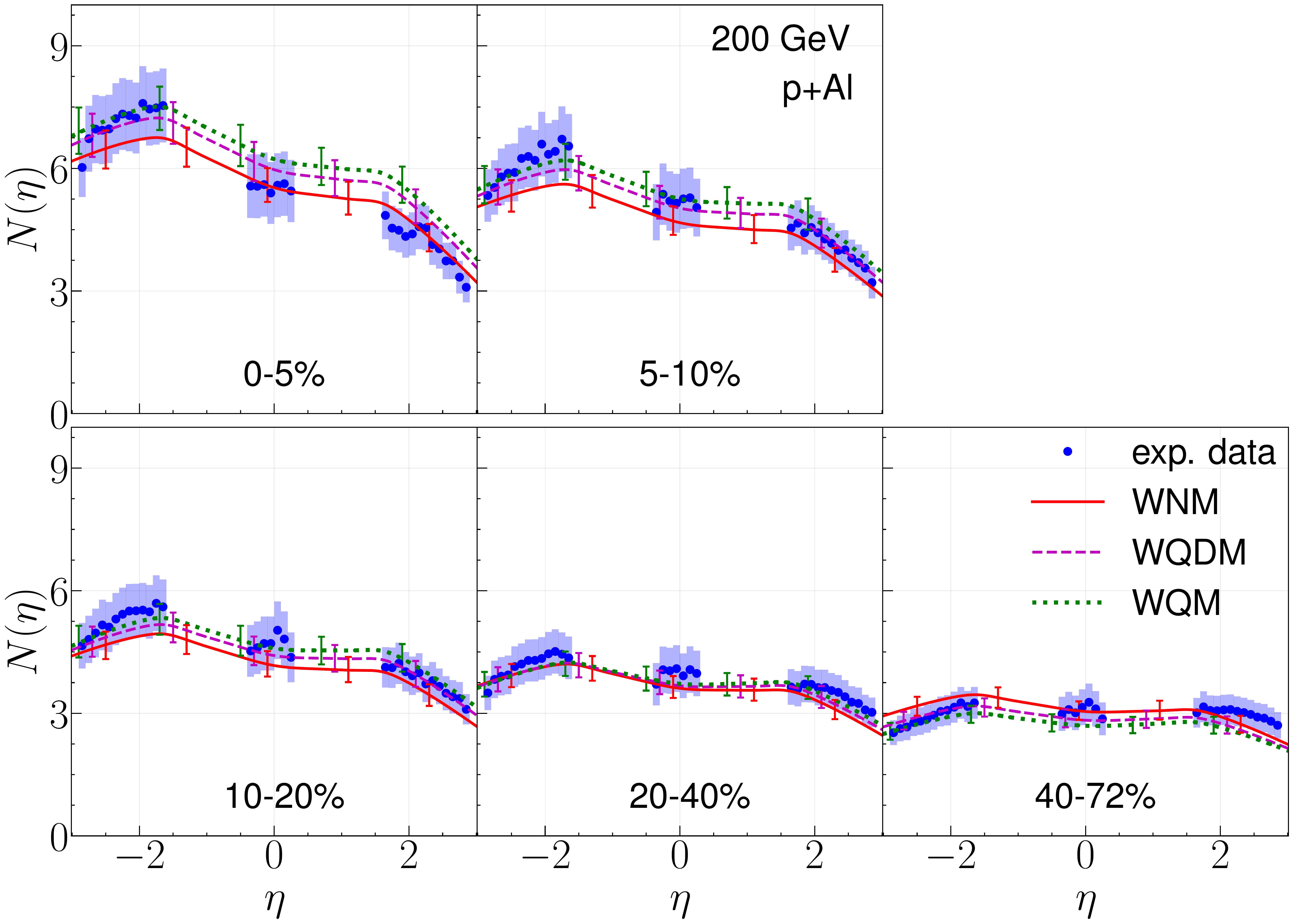}
\caption{Charged particle multiplicity distribution $N(\eta)$ as a function of pseudorapidity in the WNM (wounded nucleon model), the WQDM (wounded quark-diquark model) and the WQM (wounded quark model) in various centrality classes for p+$^{27}\!$Al collision at $\sqrt{s_{_{NN}}}=200$ GeV. Dots represent the PHENIX data \cite{Adare:2018toe}. Uncertainties are marked as bars for our simulation and as shaded areas for the experiment. Note that the measurement was carried out in the limited ranges of $\eta$.\label{fig:p-Al}}
%\end{center}
%\end{figure}
\vspace{2em}
%%%%%%%%%%%%%%%%%%%%%%%%%%%%%%%%%%%%%%%%%%%%%%%%%%%%%%%%%%%%%%%%%%%%%%%%%%%%%%%%%%%%%%%%%%
%\begin{figure}[H]
%\begin{center}
\includegraphics[scale=0.20]{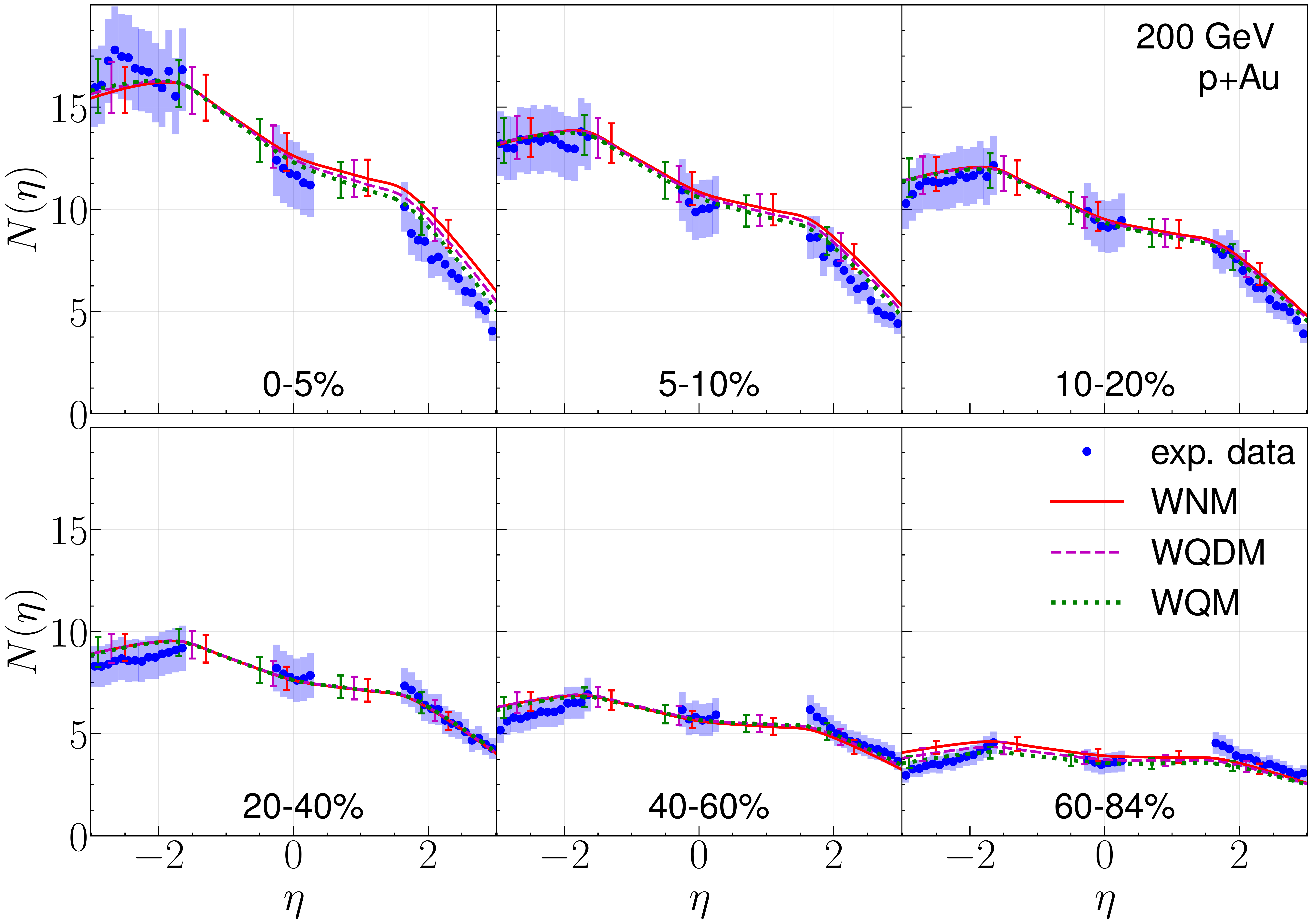}
\caption{Same as Figure \ref{fig:p-Al} but for p+$^{197}\!$Au collision.\label{fig:p-Au}}
\end{center}
\end{figure}

\begin{figure}[H]
\begin{center}
\includegraphics[scale=0.20]{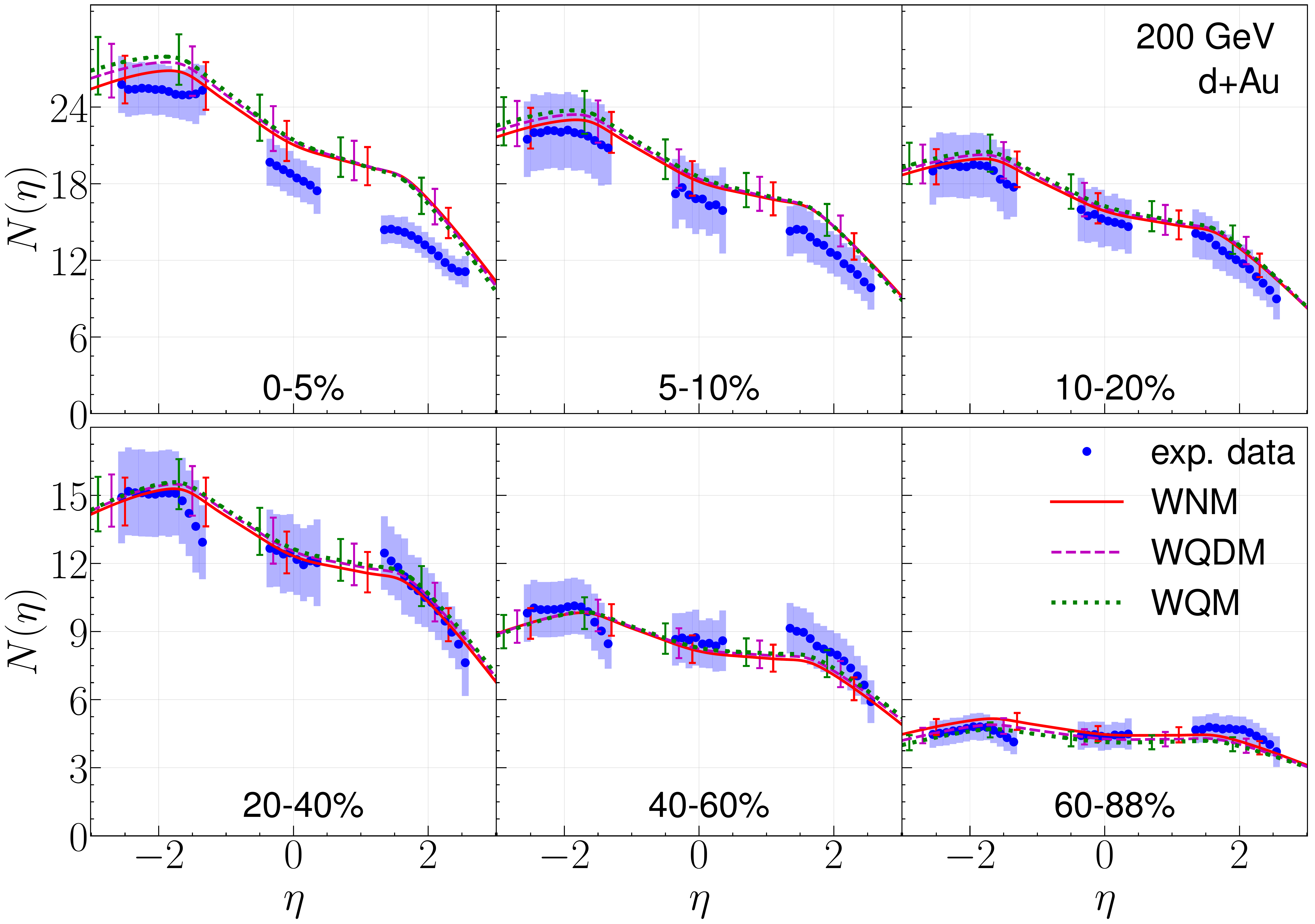}
\caption{Same as Figure \ref{fig:p-Al} but for d+$^{197}\!$Au collision.\label{fig:d-Au}}
%\end{center}
%\end{figure}
\vspace{2em}
%%%%%%%%%%%%%%%%%%%%%%%%%%%%%%%%%%%%%%%%%%%%%%%%%%%%%%%%%%%%%%%%%%%%%%%%%%%%%%%%%%%%%%%%%%
%\begin{figure}[H]
%\begin{center}
\includegraphics[scale=0.20]{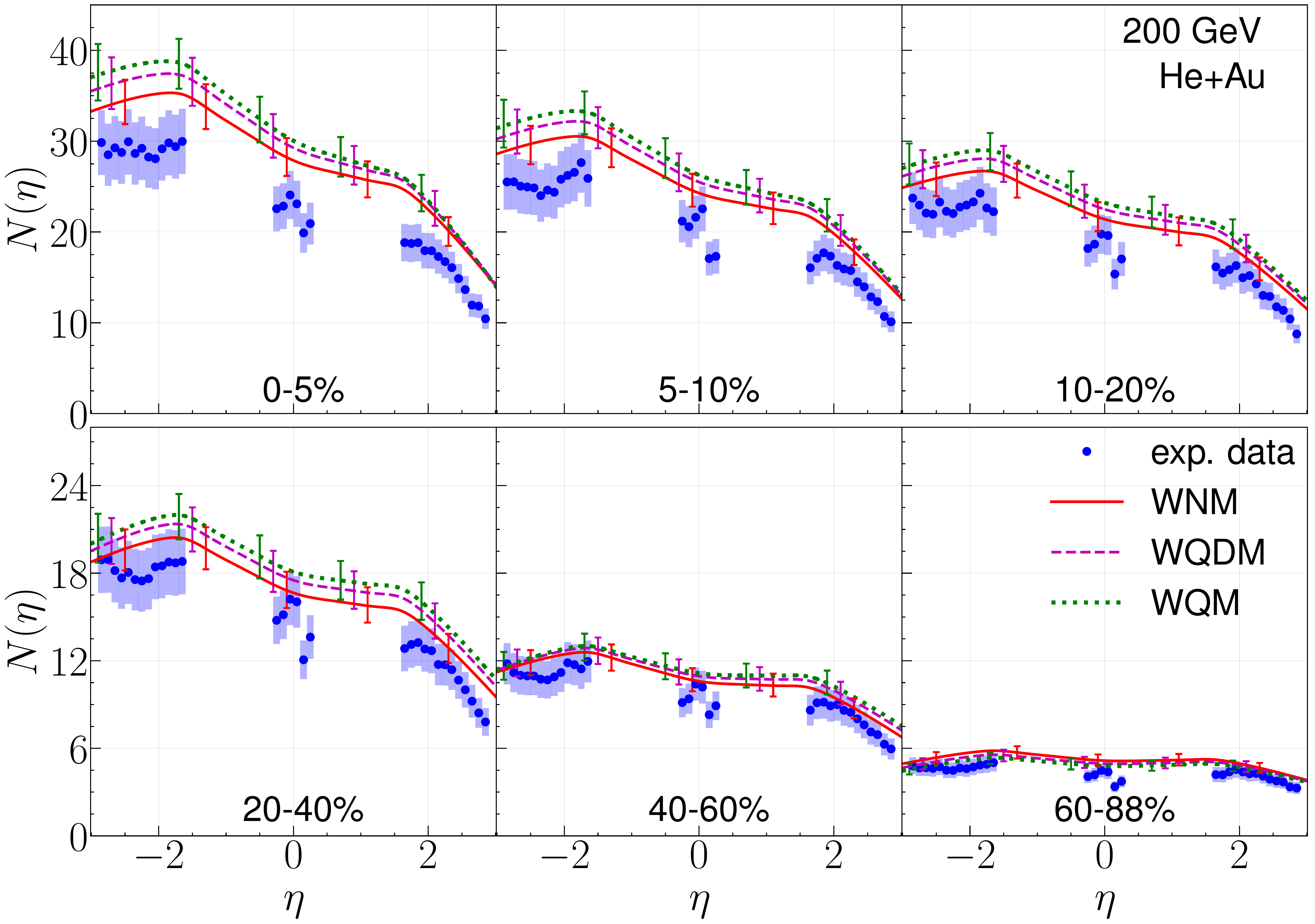}
\caption{Same as Figure \ref{fig:p-Al} but for $^3$He+$^{197}\!$Au collision.\label{fig:He-Au}}
\end{center}
\end{figure}

First, we compare our simulations in all three models with the recent PHENIX results for asymmetric collisions at $\sqrt{s_{_{NN}}} = 200\  \text{GeV}$ \cite{Adare:2018toe}. In Tables \ref{table:p-Al}, \ref{table:p-Au},\ref{table:d-Au},\ref{table:He-Au} we show the mean numbers of wounded sources in p+$^{27}\!$Al, p+$^{197}\!$Au, d+$^{197}\!$Au and $^3$He+$^{197}\!$Au collisions for different centrality classes. In Figures \ref{fig:p-Al}, \ref{fig:p-Au}, \ref{fig:d-Au}, \ref{fig:He-Au} we present the calculated pseudorapidity charged particle multiplicity distributions. We note that our predictions based on the WQM, published in Ref. \cite{Barej:2017kcw}, have already been successfully verified by PHENIX, see Ref. \cite{Adare:2018toe}.
%%%%%%%%%%%%%%%%%%%%%%%%%%%%%%%%%%%%%%%%%%%%%%%%%%%%%%%%%%%%%%%%%%%%%%%%%%%%%%%%%%%%%%%%%%
\begin{table}[H]\centering
\begin{tabular}{|R{1.5cm}|R{2cm}|R{2cm}|R{2cm}|R{2cm}|R{2cm}|R{2cm}|} \hline
 & min-bias & 0-5\% & 5-10\% & 10-20\% & 20-40\% & 40-72\% \\ \hline
WNM  & 1.00, 1.96 & 1.00, 3.85 & 1.00, 3.10 & 1.00, 2.66 & 1.00, 2.17 & 1.00, 1.67 \\ \hline
WQDM & 1.40, 2.25 & 1.85, 4.77 & 1.76, 3.81 & 1.67, 3.23 & 1.53, 2.55 & 1.31, 1.83 \\ \hline
WQM  & 1.76, 2.51 & 2.68, 5.61 & 2.51, 4.49 & 2.33, 3.78 & 2.03, 2.92 & 1.60, 1.99 \\ \hline
\end{tabular}
\caption{The mean numbers of wounded nucleons, quark-diquarks and quarks for different centrality classes in p+$^{27}\!$Al collisions at $\sqrt{s_{_{NN}}} = 200\  \text{GeV}$. The first and second numbers in each cell concern p and Al, respectively.}\label{table:p-Al}
\end{table}
%%%%%%%%%%%%%%%%%%%%%%%%%%%%%%%%%%%%%%%%%%%%%%%%%%%%%%%%%%%%%%%%%%%%%%%%%%%%%%%%%%%%%%%%%%
\begin{table}[H]\centering
\begin{tabular}{|R{1.5cm}|R{2cm}|R{2cm}|R{2cm}|R{2cm}|R{2cm}|R{2cm}|R{2cm}|} \hline
 & min-bias & 0-5\% & 5-10\% & 10-20\% & 20-40\% & 40-60\% & 60-84\% \\ \hline
WNM  & 1.00, 4.47 & 1.00, 10.07 & 1.00, 8.52 & 1.00, 7.35 & 1.00, 5.68 & 1.00, 3.93 & 1.00, 2.44 \\ \hline
WQDM & 1.66, 5.11 & 1.99, 11.84 & 1.98, 9.93 & 1.95, 8.51 & 1.89, 6.59 & 1.75, 4.57 & 1.46, 2.69 \\ \hline
WQM  & 2.30, 5.68 & 2.98, 13.40 & 2.95, 11.13 & 2.90, 9.55 & 2.77, 7.39 & 2.47, 5.08 & 1.87, 2.87 \\ \hline
\end{tabular}
\caption{Same as Table \ref{table:p-Al} but for p+$^{197\!}$Au collisions.}\label{table:p-Au}
\end{table}
%%%%%%%%%%%%%%%%%%%%%%%%%%%%%%%%%%%%%%%%%%%%%%%%%%%%%%%%%%%%%%%%%%%%%%%%%%%%%%%%%%%%%%%%%%
\begin{table}[H]\centering
\begin{tabular}{|R{1.5cm}|R{2cm}|R{2cm}|R{2cm}|R{2cm}|R{2cm}|R{2cm}|R{2cm}|} \hline
 & min-bias & 0-5\% & 5-10\% & 10-20\% & 20-40\% & 40-60\% & 60-88\% \\ \hline
WNM  & 1.59, 6.56 & 1.99, 16.48 & 1.98, 13.96 & 1.95, 11.97 & 1.86, 8.96 & 1.65, 5.49 & 1.31, 2.62 \\ \hline
WQDM & 2.65, 7.67 & 3.92, 19.75 & 3.84, 16.55 & 3.72, 14.13 & 3.39, 10.52 & 2.74, 6.40 & 1.81, 2.93 \\ \hline
WQM  & 3.68, 8.70 & 5.85, 22.68 & 5.71, 18.96 & 5.49, 16.15 & 4.89, 11.94 & 3.78, 7.25 & 2.27, 3.23 \\ \hline
\end{tabular}
\caption{Same as Table \ref{table:p-Al} but for d+$^{197\!}$Au collisions.}\label{table:d-Au}
\end{table}
%%%%%%%%%%%%%%%%%%%%%%%%%%%%%%%%%%%%%%%%%%%%%%%%%%%%%%%%%%%%%%%%%%%%%%%%%%%%%%%%%%%%%%%%%%
\begin{table}[H]\centering
\begin{tabular}{|R{1.5cm}|R{2cm}|R{2cm}|R{2cm}|R{2cm}|R{2cm}|R{2cm}|R{2cm}|} \hline
 & min-bias & 0-5\% & 5-10\% & 10-20\% & 20-40\% & 40-60\% & 60-88\% \\ \hline
WNM  & 2.30, 8.27 & 2.99, 21.46 & 2.98, 18.32 & 2.96, 15.82 & 2.86, 11.75 & 2.51, 6.78 & 1.72, 2.80 \\ \hline
WQDM & 3.82, 9.98 & 5.88, 26.59 & 5.79, 22.49 & 5.65, 19.31 & 5.21, 14.23 & 4.07, 8.07 & 2.29, 3.20 \\ \hline
WQM  & 5.30, 11.58 & 8.79, 31.21 & 8.61, 26.35 & 8.35, 22.56 & 7.52, 16.57 & 5.54, 9.30 & 2.81, 3.57 \\ \hline
\end{tabular}
\caption{Same as Table \ref{table:p-Al} but for $^3$He+$^{197\!}$Au collisions.}\label{table:He-Au}
\end{table}
%%%%%%%%%%%%%%%%%%%%%%%%%%%%%%%%%%%%%%%%%%%%%%%%%%%%%%%%%%%%%%%%%%%%%%%%%%%%%%%%%%%%%%%%%%
%%%%%%%%%%%%%%%%%%%%%%%%%%%%%%%%%%%%%%%%%%%%%%%%%%%%%%%%%%%%%%%%%%%%%%%%%%%%%%%%%%%%%%%%%%
%%%%%%%%%%%%%%%%%%%%%%%%%%%%%%%%%%%%%%%%%%%%%%%%%%%%%%%%%%%%%%%%%%%%%%%%%%%%%%%%%%%%%%%%%%
In the above figures one can observe that the differences between the models are negligible for the studied asymmetric collisions. All simulations are in quite good agreement with the results from the PHENIX collaboration \cite{Adare:2018toe}. The largest disagreement (of the order of 20\%) is for the most central $^3$He+Au collisions.

We note that $F(\eta)$ is extracted from the min-bias PHOBOS data on d+$^{197}$Au collisions and all our calculations are basically parameter-free. Consequently, one should not expect to obtain a better agreement than a few tens of percent.

%%%%%%%%%%%%%%%%%%%%%%%%%%%%%%%%%%%%%%%%%%%%%%%%%%%%%%%%%%%%%%%%%%%%%%%%%%%%%%%%%%%%%%%%%%
%%%%%%%%%%%%%%%%%%%%%%%%%%%%%%%%%%%%%%%%%%%%%%%%%%%%%%%%%%%%%%%%%%%%%%%%%%%%%%%%%%%%%%%%%%
%%%%%%%%%%%%%%%%%%%%%%%%%%%%%%%%%%%%%%%%%%%%%%%%%%%%%%%%%%%%%%%%%%%%%%%%%%%%%%%%%%%%%%%%%%

\subsection{Cu-Cu, Au-Au}
\begin{figure}[H]
\begin{center}
\includegraphics[scale=0.20]{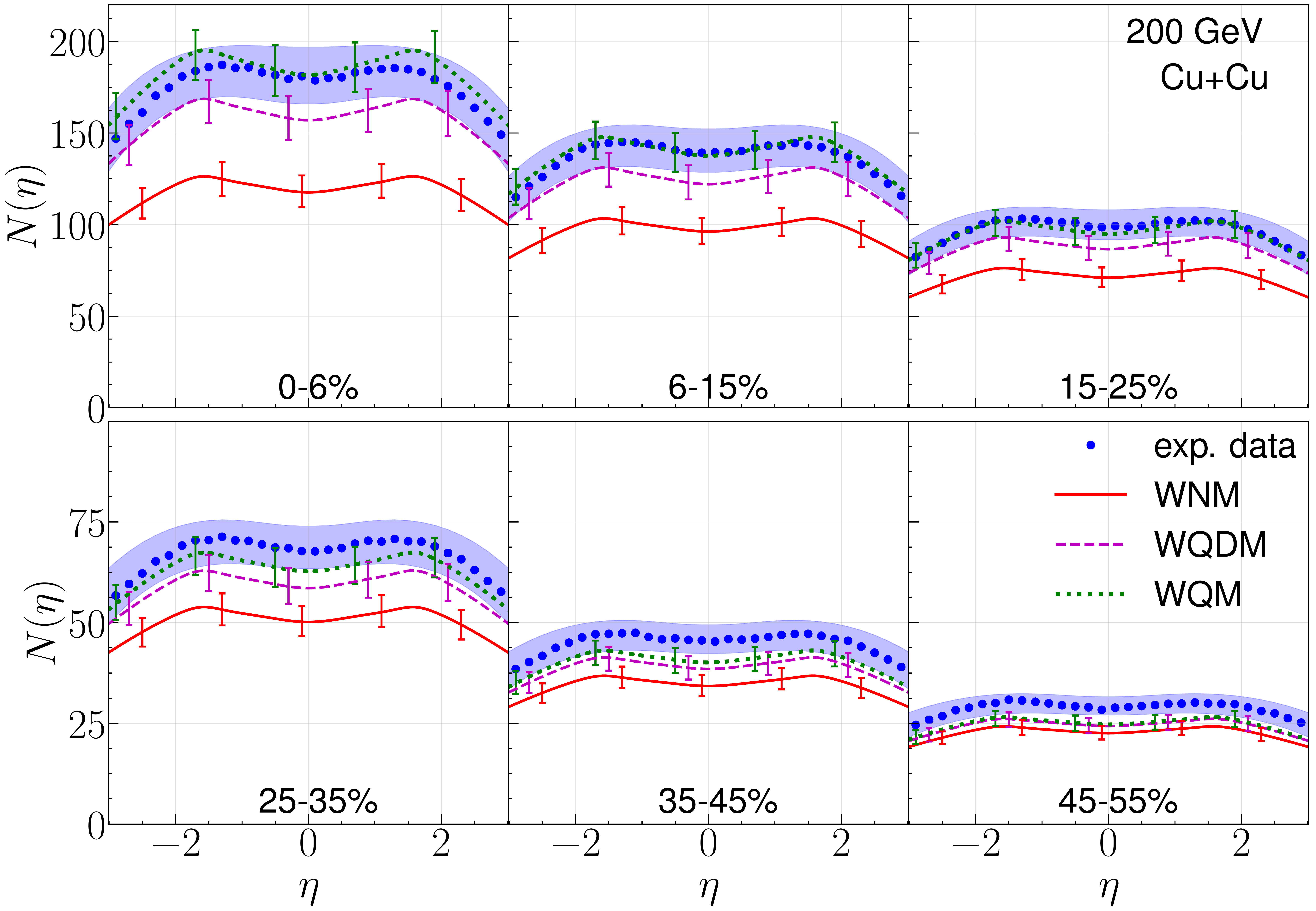}
\caption{Same as Figure \ref{fig:p-Al} but for $^{63\!}$Cu+$^{63\!}$Cu collisions. Dots represent the PHOBOS data \cite{Alver:2007aa}.}\label{fig:Cu-Cu}
%\end{center}
%\end{figure}
\vspace{2em}
%%%%%%%%%%%%%%%%%%%%%%%%%%%%%%%%%%%%%%%%%%%%%%%%%%%%%%%%%%%%%%%%%%%%%%%%%%%%%%%%%%%%%%%%%%
%\begin{figure}[H]
%\begin{center}
\includegraphics[scale=0.20]{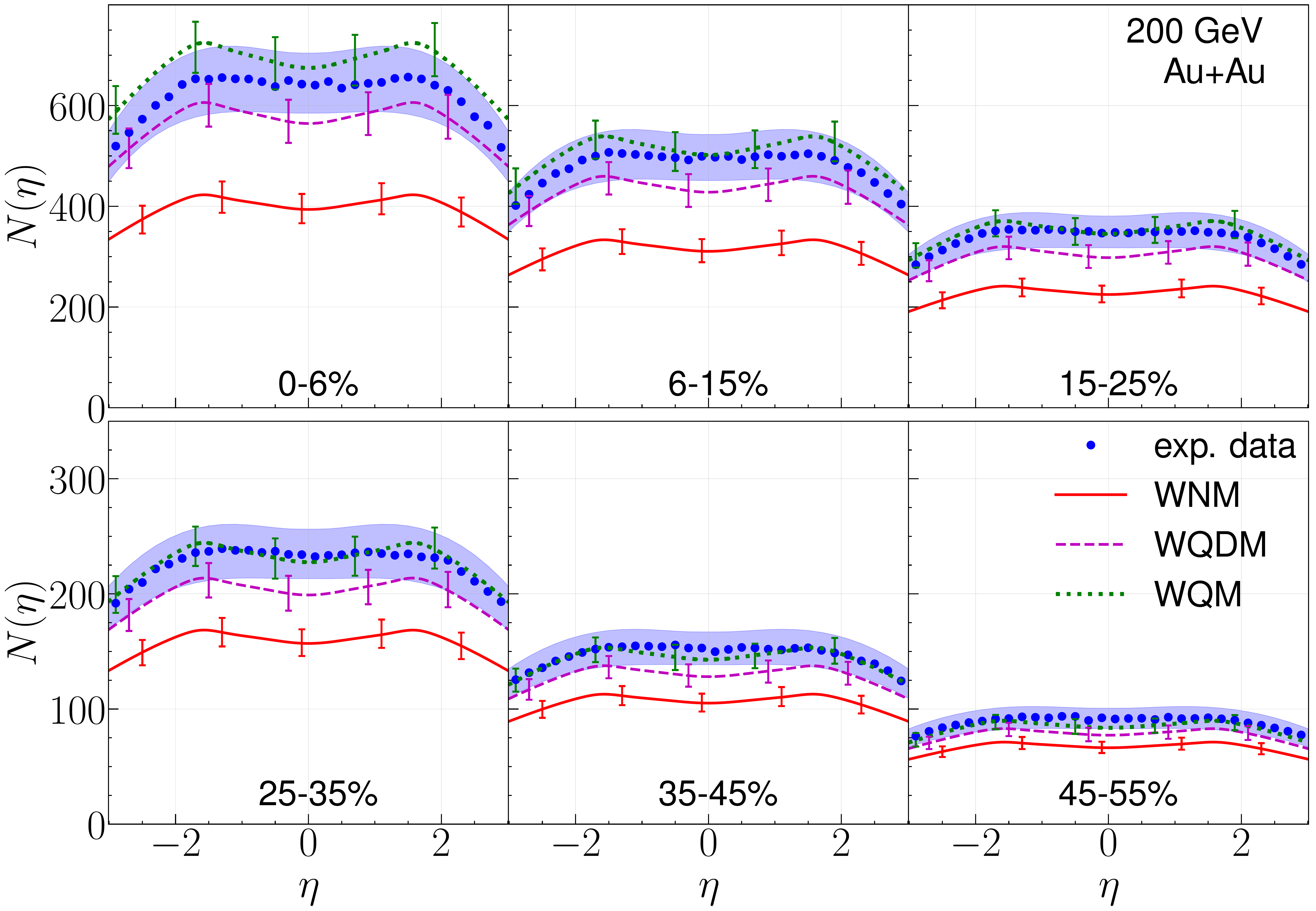}
\caption{Same as Figure \ref{fig:p-Al} but for $^{197\!}$Au+$^{197\!}$Au collisions. Dots represent the PHOBOS data \cite{Back:2002wb}.}\label{fig:Au-Au}
\end{center}
\end{figure}

Here we test all three models with the PHOBOS results on ${\D N_{\text{ch}}}/{\D \eta}$ for $^{63}$Cu+$^{63}$Cu and $^{197\!}$Au+$^{197\!}$Au collisions at $\sqrt{s_{_{NN}}}=200$ GeV. Tables \ref{table:Cu-Cu}, \ref{table:Au-Au} contain the mean numbers of wounded sources for these collisions and Figs. \ref{fig:Cu-Cu}, \ref{fig:Au-Au} demonstrate the calculated and measured pseudorapidity charged particle multiplicity distributions.

In the case of symmetric collisions, we get distributions, which differ significantly among the models. As expected, the wounded nucleon model is not valid for central collisions of heavy nuclei. On the other hand, our results from the wounded quark and quark-diquark models are in quite good agreement with the data.
%%%%%%%%%%%%%%%%%%%%%%%%%%%%%%%%%%%%%%%%%%%%%%%%%%%%%%%%%%%%%%%%%%%%%%%%%%%%%%%%%%%%%%%%%%
\begin{table}[H]\centering
\begin{tabular}{|R{1.5cm}|R{1.5cm}|R{1.5cm}|R{1.5cm}|R{1.5cm}|R{1.5cm}|R{1.5cm}|R{1.5cm}|} \hline
 & min-bias & 0-6\% & 6-15\% & 15-25\% & 25-35\% & 35-45\% & 45-55\% \\ \hline
WNM  & 16.2 & 51.6 & 42.2 & 31.2 & 22.0 & 15.0 & 9.9 \\ \hline
WQDM & 24.7 & 87.1 & 67.7 & 48.1 & 32.5 & 21.4 & 13.5 \\ \hline
WQM  & 32.5 & 121.0 & 91.6 & 63.2 & 41.8 & 26.7 & 16.5 \\ \hline
\end{tabular}
\caption{The mean numbers of wounded nucleons, quark-diquarks and quarks (per one nucleus) for different centrality classes in $^{63}$Cu+$^{63}$Cu collisions at $\sqrt{s_{_{NN}}} = 200\  \text{GeV}$.}\label{table:Cu-Cu}
\end{table}
%%%%%%%%%%%%%%%%%%%%%%%%%%%%%%%%%%%%%%%%%%%%%%%%%%%%%%%%%%%%%%%%%%%%%%%%%%%%%%%%%%%%%%%%%%
\begin{table}[H]\centering
\begin{tabular}{|R{1.5cm}|R{1.5cm}|R{1.5cm}|R{1.5cm}|R{1.5cm}|R{1.5cm}|R{1.5cm}|R{1.5cm}|} \hline
 & min-bias & 0-6\% & 6-15\% & 15-25\% & 25-35\% & 35-45\% & 45-55\% \\ \hline
WNM  & 50.2 & 172.7 & 136.3 & 98.6 & 68.9 & 46.1 & 29.1 \\ \hline
WQDM & 83.6 & 313.1 & 237.5 & 165.4 & 110.4 & 71.1 & 42.9 \\ \hline
WQM  & 115.7 & 449.1 & 334.1 & 229.8 & 151.4 & 95.0 & 55.7 \\ \hline
\end{tabular}
\caption{Same as Table \ref{table:Cu-Cu} but for $^{197\!}$Au+$^{197\!}$Au collisions.}\label{table:Au-Au}
\end{table}
%%%%%%%%%%%%%%%%%%%%%%%%%%%%%%%%%%%%%%%%%%%%%%%%%%%%%%%%%%%%%%%%%%%%%%%%%%%%%%%%%%%%%%%%%%
%%%%%%%%%%%%%%%%%%%%%%%%%%%%%%%%%%%%%%%%%%%%%%%%%%%%%%%%%%%%%%%%%%%%%%%%%%%%%%%%%%%%%%%%%%

%%%%%%%%%%%%%%%%%%%%%%%%%%%%%%%%%%%%%%%%%%%%%%%%%%%%%%%%%%%%%%%%%%%%%%%%%%%%%%%%%%%%%%%%%%
%%%%%%%%%%%%%%%%%%%%%%%%%%%%%%%%%%%%%%%%%%%%%%%%%%%%%%%%%%%%%%%%%%%%%%%%%%%%%%%%%%%%%%%%%%
%%%%%%%%%%%%%%%%%%%%%%%%%%%%%%%%%%%%%%%%%%%%%%%%%%%%%%%%%%%%%%%%%%%%%%%%%%%%%%%%%%%%%%%%%%

\subsection{Cu-Au and U-U}
\begin{figure}[H]
\begin{center}
\includegraphics[scale=0.20]{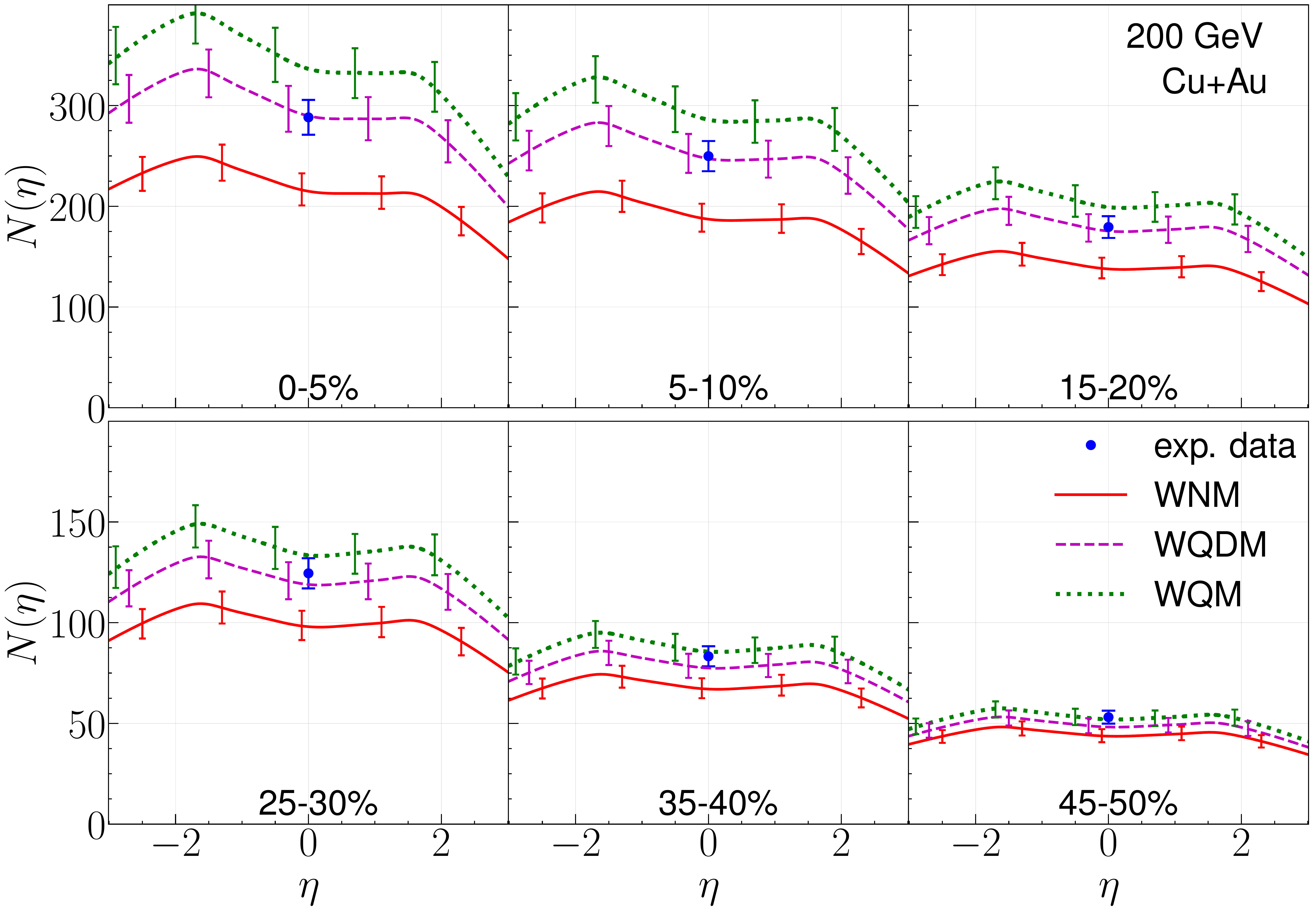}
\caption{Same as Figure \ref{fig:p-Al} but for $^{63}$Cu+$^{197\!}$Au collisions. The dots at $\eta=0$ represent the PHENIX data \cite{Adare:2015bua}.}\label{fig:Cu-Au}
\end{center}
\end{figure}

\begin{figure}[H]
\begin{center}
\includegraphics[scale=0.20]{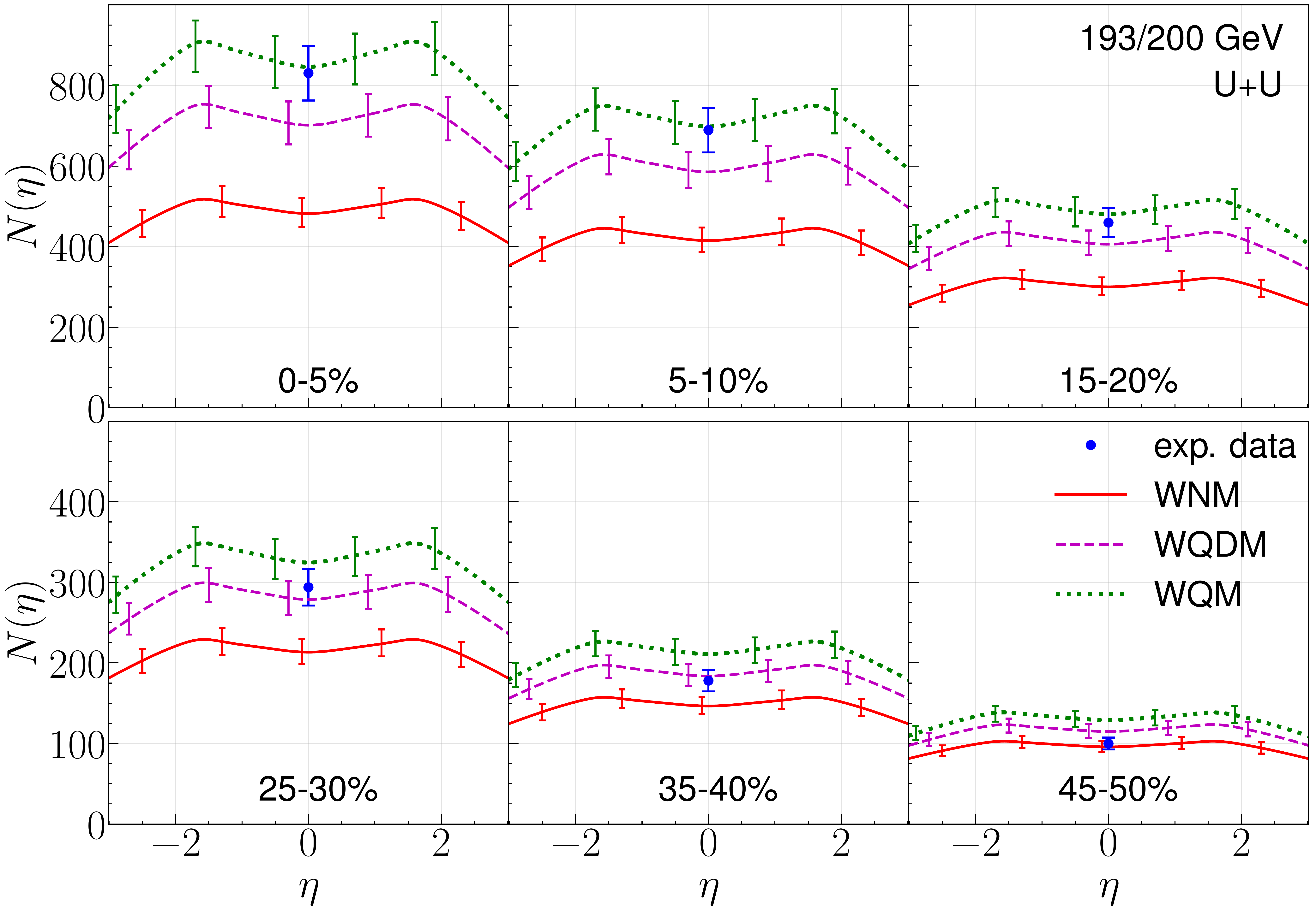}
\caption{Same as Figure \ref{fig:p-Al} but for $^{238}$U+$^{238}$U collisions. The dots 
at $\eta=0$ represent the PHENIX data for $^{238}$U+$^{238}$U at $\sqrt{s_{_{NN}}} = 193\  \text{GeV}$ \cite{Adare:2015bua}.}\label{fig:U-U}
\end{center}
\end{figure}

Next, we discuss $^{63}$Cu+$^{197}\!$Au and $^{238}$U+$^{238}$U collisions where we have very limited experimental data to compare with \cite{Adare:2015bua}. The mean numbers of wounded sources are presented in Tables \ref{table:Cu-Au}, \ref{table:U-U} and the obtained distributions ${\D N_{\text{ch}}}/{\D \eta}$ in Figures \ref{fig:Cu-Au} and \ref{fig:U-U}. For these collisions, we show only a few selected centrality bins.
%%%%%%%%%%%%%%%%%%%%%%%%%%%%%%%%%%%%%%%%%%%%%%%%%%%%%%%%%%%%%%%%%%%%%%%%%%%%%%%%%%%%%%%%%%
\begin{table}[H]\centering
\begin{tabular}{|R{1.5cm}|R{2cm}|R{2cm}|R{2cm}|R{2cm}|R{2cm}|R{2cm}|R{2cm}|} \hline
& min-bias & 0-5\% & 5-10\% & 15-20\% & 25-30\% & 35-40\% & 45-50\% \\ \hline
WNM  & 22.1, 34.9 & 61.1, 127.5 & 57.8, 106.5 & 47.2, 73.7 & 35.4, 50.6 & 25.1, 33.8 & 16.7, 21.6 \\ \hline
WQDM & 36.4, 52.6 & 115.6, 205.9 & 105.5, 168.9 & 80.5, 114.2 & 56.9, 75.1 & 38.0, 47.9 & 24.1, 29.4 \\ \hline
WQM  & 50.7, 69.9 & 169.3, 278.6 & 152.2, 228.5 & 112.9, 152.3 & 78.2, 99.4 & 51.3, 62.7 & 31.7, 37.6 \\ \hline
\end{tabular}
\caption{Same as Table \ref{table:p-Al} but for $^{63}$Cu+$^{197\!}$Au collisions.}\label{table:Cu-Au}
\end{table}
%%%%%%%%%%%%%%%%%%%%%%%%%%%%%%%%%%%%%%%%%%%%%%%%%%%%%%%%%%%%%%%%%%%%%%%%%%%%%%%%%%%%%%%%%%
\begin{table}[H]\centering
\begin{tabular}{|R{1.5cm}|R{1.5cm}|R{1.5cm}|R{1.5cm}|R{1.5cm}|R{1.5cm}|R{1.5cm}|R{1.5cm}|} \hline
& min-bias & 0-5\% & 5-10\% & 15-20\% & 25-30\% & 35-40\% & 45-50\% \\ \hline
WNM  & 62.0 & 211.5 & 182.1 & 131.6 & 93.6 & 64.3 & 42.0 \\ \hline
WQDM & 104.5 & 389.1 & 324.8 & 225.2 & 154.6 & 102.0 & 63.7 \\ \hline
WQM  & 146.8 & 563.4 & 464.6 & 319.7 & 216.1 & 140.5 & 85.9 \\ \hline
\end{tabular}
\caption{Same as Table \ref{table:Cu-Cu} but for $^{238}$U+$^{238}$U collisions.}\label{table:U-U}
\end{table}
%%%%%%%%%%%%%%%%%%%%%%%%%%%%%%%%%%%%%%%%%%%%%%%%%%%%%%%%%%%%%%%%%%%%%%%%%%%%%%%%%%%%%%%%%%
%%%%%%%%%%%%%%%%%%%%%%%%%%%%%%%%%%%%%%%%%%%%%%%%%%%%%%%%%%%%%%%%%%%%%%%%%%%%%%%%%%%%%%%%%%
%%%%%%%%%%%%%%%%%%%%%%%%%%%%%%%%%%%%%%%%%%%%%%%%%%%%%%%%%%%%%%%%%%%%%%%%%%%%%%%%%%%%%%%%%%

Results for $^{63}$Cu+$^{197}\!$Au and $^{238}$U+$^{238}$U again indicate that the WQDM and the WQM are in acceptable agreement with the experimental data. We note that in the case of $^{238}$U+$^{238}$U collisions $\sqrt{s_{_{NN}}} = 193$ GeV and in our simulations we used the wounded constituent emission functions extracted at $\sqrt{s_{_{NN}}} = 200$ GeV. It is rather obvious that this small difference in energy is negligible at our accuracy level.

\subsection{p+p}
\begin{figure}[H]
\begin{center}
\includegraphics[scale=0.16]{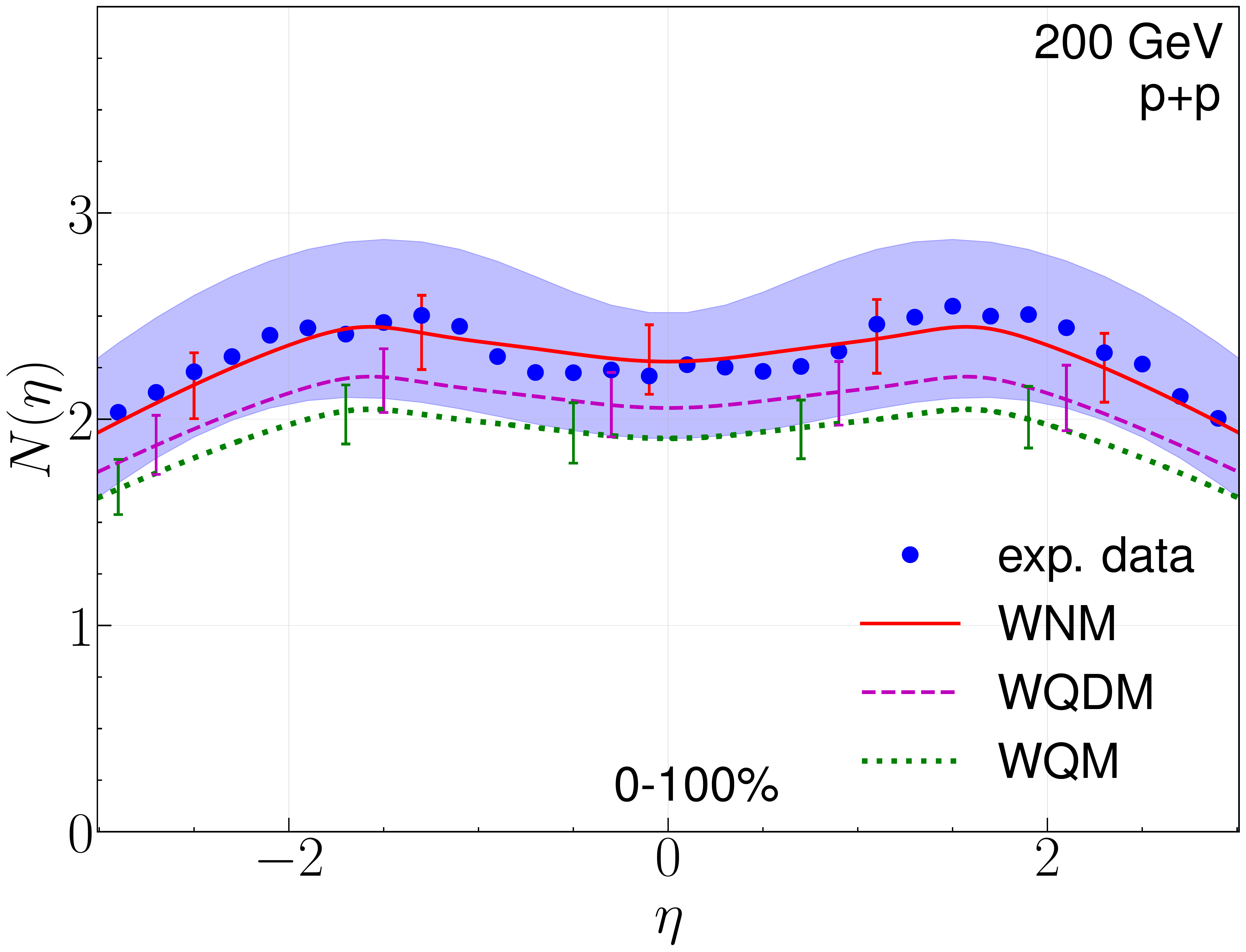}
\caption{Same as Figure \ref{fig:p-Al} but for proton-proton collisions. Dots represent the PHOBOS data \cite{Alver:2010ck}.}\label{fig:p-p}
\end{center}
\end{figure}

Lastly, we present our calculations for proton-proton collisions. The obtained mean numbers of wounded sources, per one wounded proton, are 1.00, 1.14 and 1.27 for the wounded nucleon, quark-diquark and quark models, respectively. Figure \ref{fig:p-p} shows the calculated and measured by PHOBOS pseudorapidity charged particle multiplicity distributions.
%%%%%%%%%%%%%%%%%%%%%%%%%%%%%%%%%%%%%%%%%%%%%%%%%%%%%%%%%%%%%%%%%%%%%%%%%%%%%%%%%%%%%%%%%%
%%%%%%%%%%%%%%%%%%%%%%%%%%%%%%%%%%%%%%%%%%%%%%%%%%%%%%%%%%%%%%%%%%%%%%%%%%%%%%%%%%%%%%%%%%

As seen in Figure \ref{fig:p-p}, in p+p collisions all models agree (within uncertainties) with the data.

%%%%%%%%%%%%%%%%%%%%%%%%%%%%%%%%%%%%%%%%%%%%%%%%%%%%%%%%%%%%%%%%%%%%%%%%%%%%%%%%%%%%%%%%%%
%%%%%%%%%%%%%%%%%%%%%%%%%%%%%%%%%%%%%%%%%%%%%%%%%%%%%%%%%%%%%%%%%%%%%%%%%%%%%%%%%%%%%%%%%%
%%%%%%%%%%%%%%%%%%%%%%%%%%%%%%%%%%%%%%%%%%%%%%%%%%%%%%%%%%%%%%%%%%%%%%%%%%%%%%%%%%%%%%%%%%

\section{Conclusions}
Our conclusions can be formulated as follows:

\begin{enumerate}[label=(\roman*)]
  \item Using the wounded nucleon, quark and quark-diquark models we calculated three different wounded source emission functions based on the PHOBOS min-bias d+$^{197}\!$Au data at $\sqrt{s_{_{NN}}}=200\ \text{GeV}$ \cite{Back:2004mr}.  
   
  \item Using the min-bias emission functions we calculated the pseudorapidity charged particle multiplicity distributions for p+p, p+$^{27}\!$Al, p+$^{197}\!$Au, d+$^{197}\!$Au, $^3$He+$^{197}\!$Au, $^{63}$Cu+$^{63}$Cu, $^{63}$Cu+$^{197}\!$Au, $^{197}\!$Au+$^{197}\!$Au and $^{238}$U+$^{238}$U collisions at $\sqrt{s_{_{NN}}}=200\ \text{GeV}$ in various centrality classes. We note that once the min-bias emission function is known, the rest of our calculation is essentially free of any adjustable parameters. 

  \item All three considered models are in good agreement with the PHENIX data for highly asymmetric collisions. This is not surprising since for these collisions there is only a limited number of nucleons that collide multiple times.
  
  \item Results for symmetric collisions of heavy nuclei such as $^{63}$Cu+$^{63}\!$Cu and $^{197}\!$Au+$^{197}\!$Au differ significantly among the models. Both wounded quark and quark-diquark models are in quite good agreement with the PHOBOS data. As expected, the wounded nucleon model underpredicts the data, except for very peripheral ones. For p+p interactions, all three models are acceptable.
  
  \item The wounded quark and quark-diquark models are in quite good agreement with the PHENIX data \cite{Adare:2015bua} on $^{63}$Cu+$^{197}\!$Au and $^{238}$U+$^{238}$U collisions.
%at $\sqrt{s_{_{NN}}}=200\ \text{GeV}$. 
Here we could compare only at $\eta=0$ and hopefully, our predictions in the wider range of $\eta$ will be verified experimentally.
  
  \item Currently, we are working on extending these models to include the fragmentation regions. This requires including the unwounded sources (within wounded nucleons) of charged particles \cite{abab}. Moreover, it would be also desired to test the discussed models at various collision energies.
  
  \item It would be interesting to confront the recent PHENIX data \cite{Adare:2018toe} with different (and much more sophisticated) models of particle production, such as, e.g., \cite{Jeon:1997bp,Wang:1991hta,Bleicher:1999xi,Lin:2004en,Pierog:2009zt}. In particular, it would be important to test the color glass condensate inspired models \cite{Iancu:2002xk,Gelis:2010nm,Kharzeev:2002ei}.
  
\end{enumerate}

 \begin{acknowledgements}
 This work was partially supported by the Faculty of Physics and Applied Computer 
 Science AGH UST statutory tasks No. 11.11.220.01/1 within subsidy of Ministry of Science 
 and Higher Education.
 \end{acknowledgements}

\appendix
\section{}

\begin{figure}[H]
\begin{center}
\subfloat{{\includegraphics[scale=0.25]{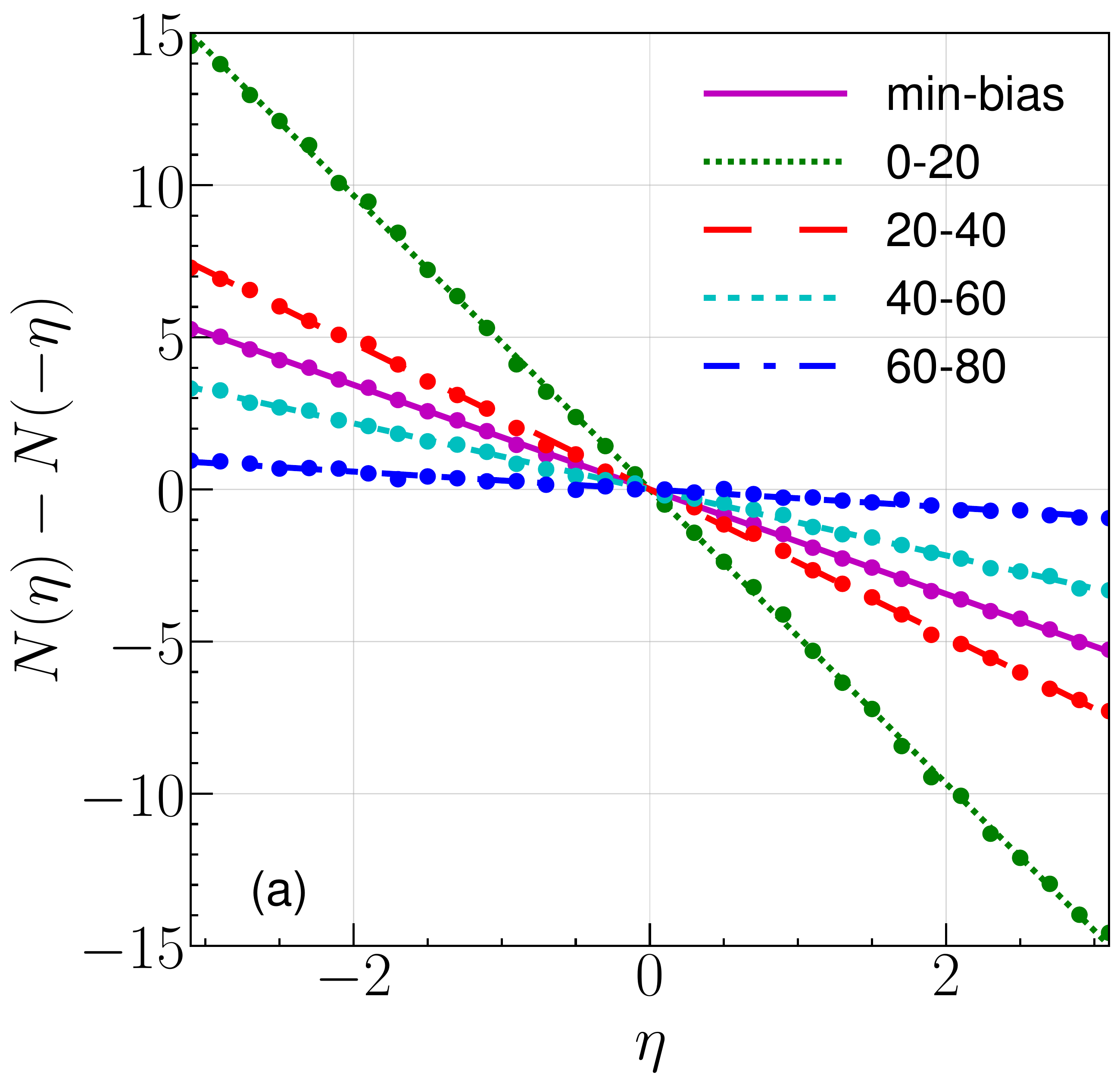}}}%
\hspace{1.0cm}
\subfloat{{\includegraphics[scale=0.25]{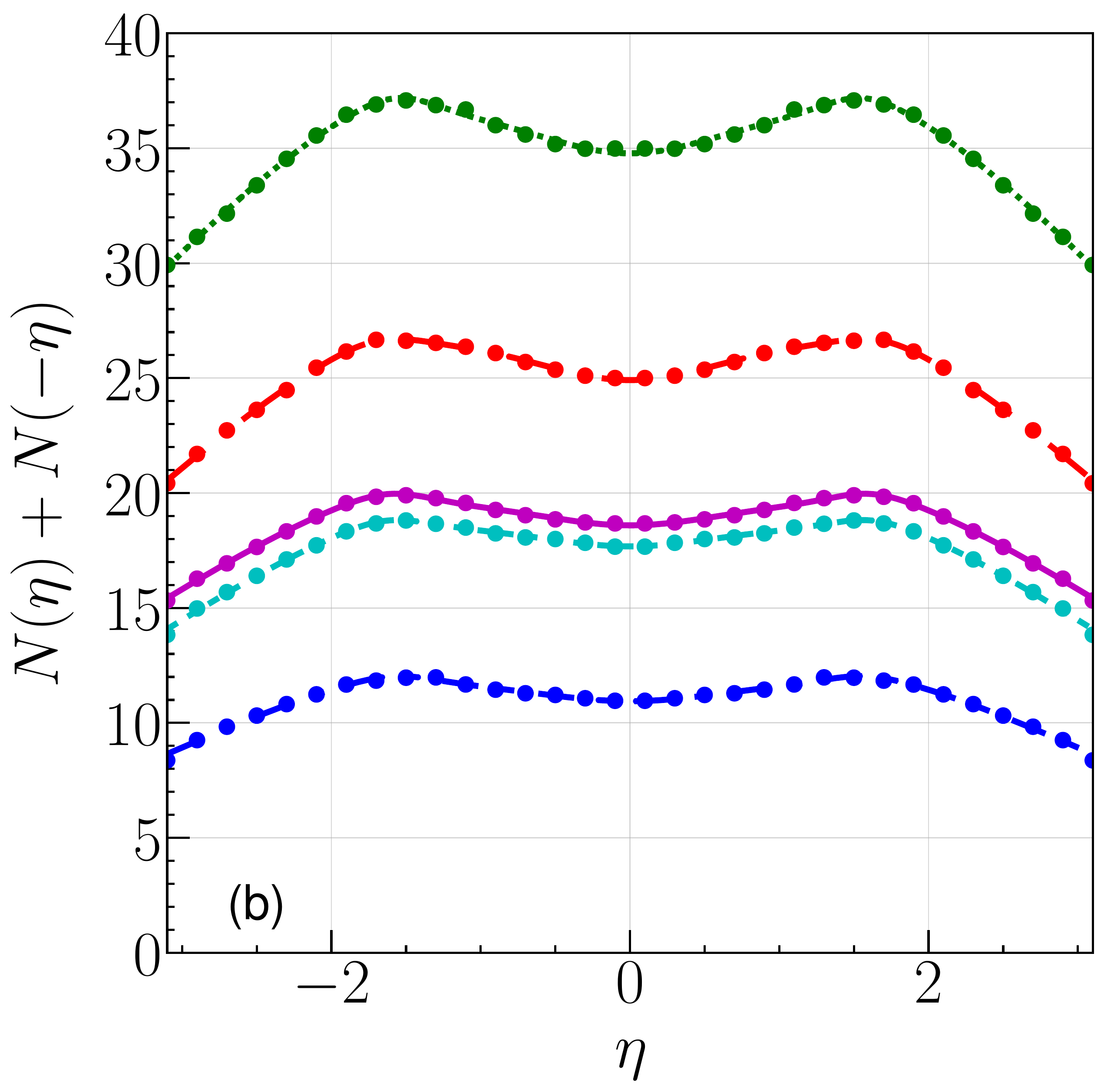}}}%
\caption{Fits (lines) to the antisymmetrized (a) and the symmetrized (b) PHOBOS data (dots) for d+$^{197\!}$Au collisions in $|\eta|<3$ at $\sqrt{s_{_{NN}}}=200$ GeV \cite{Back:2004mr}.}\label{fig:Nmp}
\end{center}
\end{figure}

In order to make the experimental input more convenient, we made fits to the PHOBOS data for d+$^{197\!}$Au collisions \cite{Back:2004mr}. Two independent functions are fitted to the symmetrized and antisymmetrized PHOBOS data. As seen in Fig. \ref{fig:Nmp}(a), the antisymmetrized case, $N^-(\eta)\equiv N(\eta)-N(-\eta)$, in a considered pseudorapidity range $\eta \in [-3,3]$, is clearly a linear function of $\eta$, that is $N^-(\eta)=c\eta$. In the symmetrized case, $N^+(\eta)\equiv N(\eta)+N(-\eta)$, the resulting points are naturally resembling a normal distribution in rapidity transformed to pseudorapidity, see Fig. \ref{fig:Nmp}(b). Obviously, $N_{\text{fit}}(\eta)=\big(N^+(\eta)+N^-(\eta)\big)/{2}$ resulting in
\begin{equation}\label{eq2}
N_{\text{fit}}(\eta)=\frac{1}{2}\Bigg[A\exp\bigg(\frac{-y^2(\eta)}{2\sigma_1^2}\bigg)\frac{T\cosh\eta}{\sqrt{1+T^2\sinh^2\eta}}+c\eta\Bigg]\,,
\end{equation}
where $A$, $\sigma_1$, $c$ are fit parameters, $y$ is rapidity and satisfies $y=\ln\left(\sqrt{1+T^2\sinh^2\eta}+T\sinh\eta\right)$, $T$ is a ratio of transverse momentum to transverse mass and is extracted from the fit. 
%%%%%%%%%%%%%%%%%%%%%%%%%%%%%%%%%%%%%%%%%%%%%%%%%%%%%%%%%%%%%%%%%%%%%%%%%%%%%%%%%%%%%%%%%%
%%%%%%%%%%%%%%%%%%%%%%%%%%%%%%%%%%%%%%%%%%%%%%%%%%%%%%%%%%%%%%%%%%%%%%%%%%%%%%%%%%%%%%%%%%
To obtain slightly better fits around $\eta=0$ for the symmetric case, as shown in Fig. \ref{fig:Nmp}(b), we multiplied the first term in Eq. (\ref{eq2}) by $\big(1-\alpha\exp\!\big(\frac{-y^6(\eta)}{2\sigma_2^{\,2}}\big)\,\big)$, where $\alpha,\sigma_2$ are new fit parameters with $\alpha$ much smaller than 1. The results of this fit differ from that of Eq. (\ref{eq2}) by about one percent (mostly in $|\eta|<1.5$ region).

\end{document}